\documentclass[]{iscram}

\usepackage[utf8]{inputenc}

\usepackage{lipsum}
\usepackage{tikz}
\usepackage{fancyvrb}
\usepackage{listings}
\usepackage{enumitem}
\usepackage{tcolorbox}
\usepackage{multicol}
\usepackage{siunitx}
\usepackage{xpatch}
\usepackage{subfigure}
\usepackage{multirow}
\usepackage{flushend}
\usepackage{makecell}
\usepackage{hyperref}
\lstdefinestyle{common}{
  xleftmargin=.5em,
  xrightmargin=.5em,
  frame=single,framesep=.5em,framerule=0pt,
  fancyvrb=true,
  basicstyle=\ttfamily,
  keywordstyle=\color{cyan!50!blue!75!black}\bfseries,
  commentstyle=\color{red!50!black}\itshape,
  stringstyle=\ttfamily\color{green!50!black},
  numbers=none,
  showspaces=false,
  showstringspaces=false,
  fontadjust=true,
  keepspaces=true,
  flexiblecolumns=true,
  emphstyle=\color{red},
}
\lstdefinestyle{TeX}{
  style=common,
  backgroundcolor=\color{blue!5},
  aboveskip=5pt,
  belowskip=5pt,
  language=[LaTeX]TeX,
  moretexcs={
    abstract, addbibresource, iscramset, keywords, mainmatter,
    maketitle, printbibliography, subsection, subsubsection, url,
    urldef, href, includegraphics, ldots, parencite, citeauthor,
    citeyear, citetitle, midrule, toprule, bottomrule
  },
  fancyvrb=true,
}
\lstdefinestyle{console}{
  style=common,
  backgroundcolor=\color{gray!10},
  aboveskip=5pt,
  belowskip=5pt,
}

\newlist{options}{description}{1}
\setlist[options]{%
  beginpenalty=10000,%
  itemsep=.5\parskip plus .3\parskip minus .2\parskip,
  parsep=.5\parskip plus .3\parskip minus .2\parskip,
  topsep=.5\parskip plus .3\parskip minus .2\parskip,
  partopsep=.5\parskip plus .3\parskip minus .2\parskip,
  style=nextline,labelindent=1em,%
  font=\normalfont\ttfamily}

\colorlet{macro color}{cyan!50!blue!75!black}
\colorlet{option color}{red!50!black}
\colorlet{generic color}{green!40!black}

\newtcolorbox{pseudoTeX}{colback=blue!5,colframe=blue!5,before=\nobreak}
\let\LaTeXorig\LaTeX
\renewcommand\LaTeX{\bgroup\fontfamily{lmr}\selectfont\upshape\LaTeXorig\egroup}

\urldef{\sitewebgind}\url{http://gind.mines-albi.fr}
\urldef{\sitewebminesalbi}\url{http://www.mines-albi.fr}
\urldef{\gmailpaulgaborit}\url{email adress}
\urldef{\jdmail}\url{...@example.com}
\urldef{\sitex}\url{www.example.com}

\urldef{\umail}\url{usa3@pitt.edu}
\urldef{\vmail}\url{viz@pitt.edu}
\urldef{\mmail}\url{prashk@pitt.edu}

\iscramset{
  CoRe Paper 2023={AI for Crisis Management},
  title={
   Localization of Events Using Neural Networks in Twitter Data\\
  },
  short title={Event localization},
  author={
    full name=Usman Anjum,
    affiliation={
      \textit{Dept. of Informatics and Networked Systems}\\ University of Pittsburgh, Pittsburgh, USA \\
     \href{usa3@pitt.edu}{\umail}
    },
  },
  author={
    full name=Vladimir Zadorozhny,
    affiliation={
      \textit{Dept. of Informatics and Networked Systems}\\ University of Pittsburgh, Pittsburgh, USA \\
      \href{viz@pitt.edu}{\vmail}
    },
  },
  author={
    full name= Prashant Krishnamurthy,
    affiliation={
      \textit{Dept. of Informatics and Networked Systems}\\ University of Pittsburgh, Pittsburgh, USA \\
      \href{prashk@pitt.edu}{\mmail}
    },
  },
}

\usepackage{tikzpagenodes}
\usetikzlibrary{fit}

\begin{document}

\maketitle

\makeatletter
%{\centering\large\iscram@version{1.0}\\\iscram@date\par}
\makeatother

\begin{abstract}
Twitter (one example of microblogging) is widely being used by researchers to understand human behavior, specifically how people behave when a significant event occurs and how it changes user microblogging patterns. The changing microblogging behavior can reveal patterns that can help in detecting real-world events. However, the Twitter data that is available has limitations, such as, it is incomplete and noisy and the samples are irregular. In this paper we create a model, called \textit{Twitter Behavior Agent-Based Model (TBAM)} to simulate Twitter pattern and behavior using Agent-Based Modeling (ABM). The generated data from ABM simulations can be used in place or to complement the real-world data toward improving the accuracy of event detection. We confirm the validity of our model by finding the cross-correlation between the real data collected from Twitter and the data generated using TBAM.
\end{abstract}


\section{Introduction}

The widespread use of microblogging services, such as Twitter, which generate immense content has resulted in considerable research focusing on utilizing their counts and semantic content for many different practical applications. For example, researchers can use microblogging data to gain insight into events and how people behave when an event occurs. The change in microblogging behavior when an event occurs creates patterns that can aid in detecting events. Detecting an event is important as it allows local authorities to both respond to the event and inform the public in a timely manner \parencite{ozdikis2017survey}.

An event is defined as a real-world one-time occurrence that generates the interest of people and is based on specific spatial and temporal properties \parencite{atefeh2015survey,ozdikis2017survey}. Events have been classified as unexpected or expected \parencite{atefeh2015survey,ozdikis2017survey}. Unexpected events are rare or at least infrequent occurrences that are unpredictable, unidentified, unscheduled or unknown. Prior knowledge about event type, time and location may not be readily available until well after the event has occurred.

The purpose of this paper is to create a novel approach, called \textit{Twitter Behavior Agent-Based Model (TBAM)}, that can simulate microblogging behavior in an event. The necessity for creating the model is due to the limitations researchers have with the real world Twitter data. Such data are scarce and unreliable in terms of the delivery, knowledge of ground truth, and information content. Before using the real world content, further complicated processing of the data would be required so that it is suitable for event detection. The data generated from simulations can be used to understand patterns or to enrich this \textit{underdeveloped} data. 

We define underdeveloped data under two dimensions. These two dimensions are \textit{reliability} and \textit{delivery slate}. Data can have high or low reliability and data delivery can be regular or sporadic. We can consider medical data from instruments (e.g. EKG data) to have high reliability and regular delivery slate. On the other hand, Twitter data has low reliability and low delivery slate.

Twitter content involves humans as source of the data which means that the data could intentionally or unintentionally be distorted and further only provides subjective information mixed with personal emotion. For example, microblogging data usually does not contain complete spatial information (like latitude and longitude) that is essential for accurate event detection. Use of location anonymization techniques for privacy preservation makes the latitude and longitude (\textit{geotags}) not readily available. Moreover, the data available is typically aggregated in space which reduces the event detection accuracy. Alternatively, researchers have used location names found in the message (also called place name) and user location found in their profiles to localize an event. But again, this information is not reliable as users may use multiple locations or may be slow in updating location information. This means that the location in the profile and the user location need not be consistent. Users may include incorrect location information which further reduces the data reliability \parencite{atefeh2015survey}. Twitter messages are short in length and can contain ambiguous words making it hard to obtain correct semantic information from the messages. Hence, Twitter data has low reliability. Another example of data with low reliability is mis-information data like fake news data \parencite{mcnair2017fake} because fake news data also involves humans as data sources.

\begin{figure}
  \centering
  \includegraphics[width=0.6\textwidth]{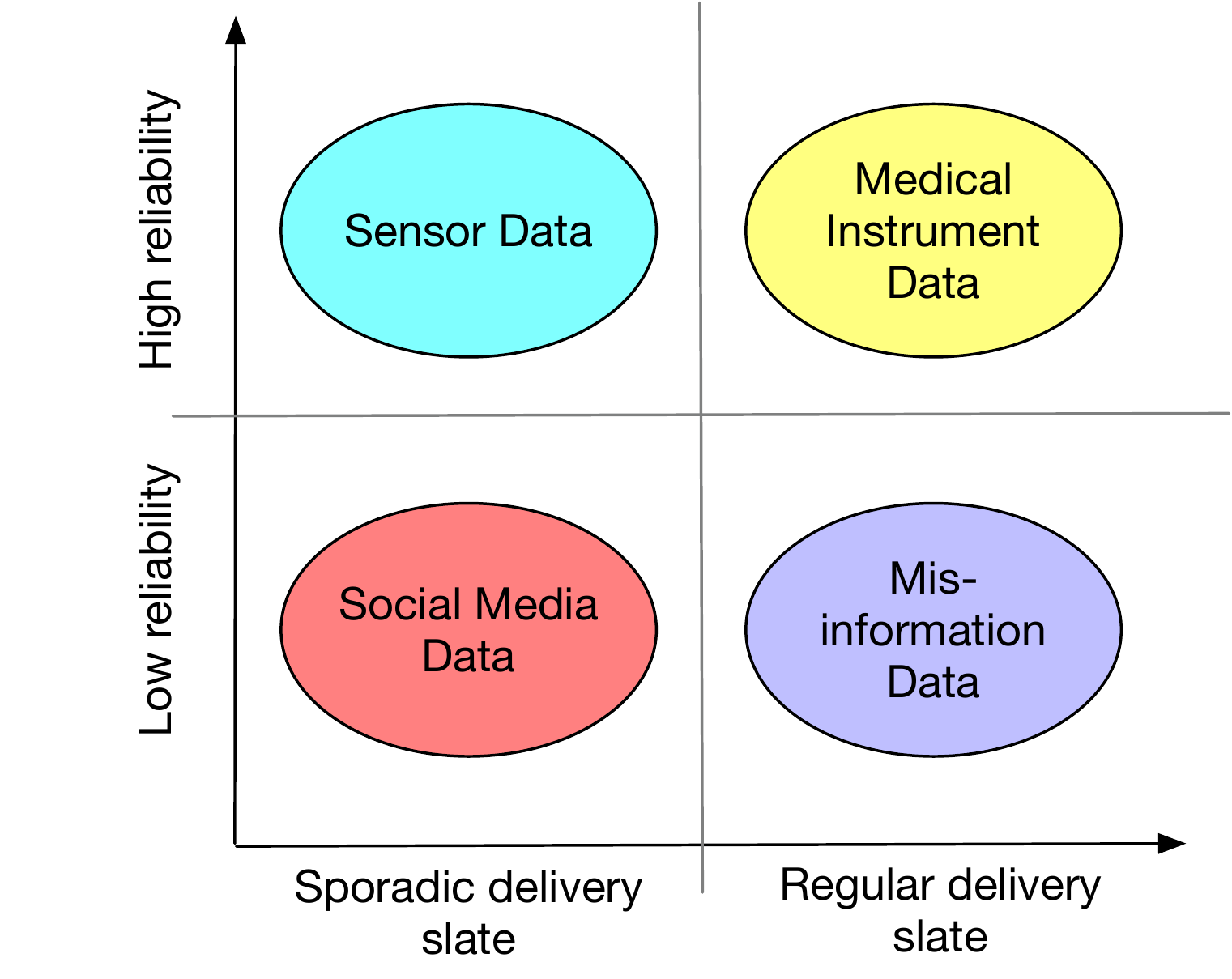} 
  \caption{Dimensions of Data}
  \label{fig:data_domain}
\end{figure}

The sporadic delivery slate of Twitter data is because there is no control over the delivery frequency and not all users are sending out event related data and even if they do, it may not be about an event and users may microblog only when it is convenient or of their interest. Factors like number of users who actively microblog, time of day, population density, significance of the microblog or event, etc. influence how frequently people may send out standard and event-related microblogs. Sensor data is another example of data with potentially sporadic delivery slates because of the battery life, duty cycle, random triggers, etc. For example, many sensors with limited battery life that cannot be replaced for long periods of time generate data only when events are detected and sometimes with long duty cycles. Other sensors only send out data when there is an external trigger (e.g. motion detection sensors only are triggered and send data when motion is detected). Thus, sensors may have unreliable delivery slate. Figure \ref{fig:data_domain} shows the placement of different data types according to their reliability and delivery slate.

A consequence of the underdeveloped nature of data is that there might not be granular data or location information. These limitations reduce event detection accuracy. Data generated through models could not only be used as a replacement to real data to understand event change patterns but also complement and enrich the real data by providing information that the real data may not contain. As one example, the generated data may be used to train machine learning models to find event signatures. These machine learning models can then be applied to real data from Twitter to detect events. To create TBAM we use agent-based models (ABM). An ABM implements a top-down modeling approach where we can set different parameters that change the generation and distribution of tweets. An alternative approach for modelling could be through machine-learning, like generative adversarial networks (GAN) which would work as a bottom-up approach. In the bottom-up approach, instead of using parameters to generate data, the real data is directly used to train GAN and generate synthetic data. The limitation of the latter is the lack of explain-ability of the synthetic data.

\begin{figure}[htp]
  \centering
  \includegraphics[width=0.80\textwidth]{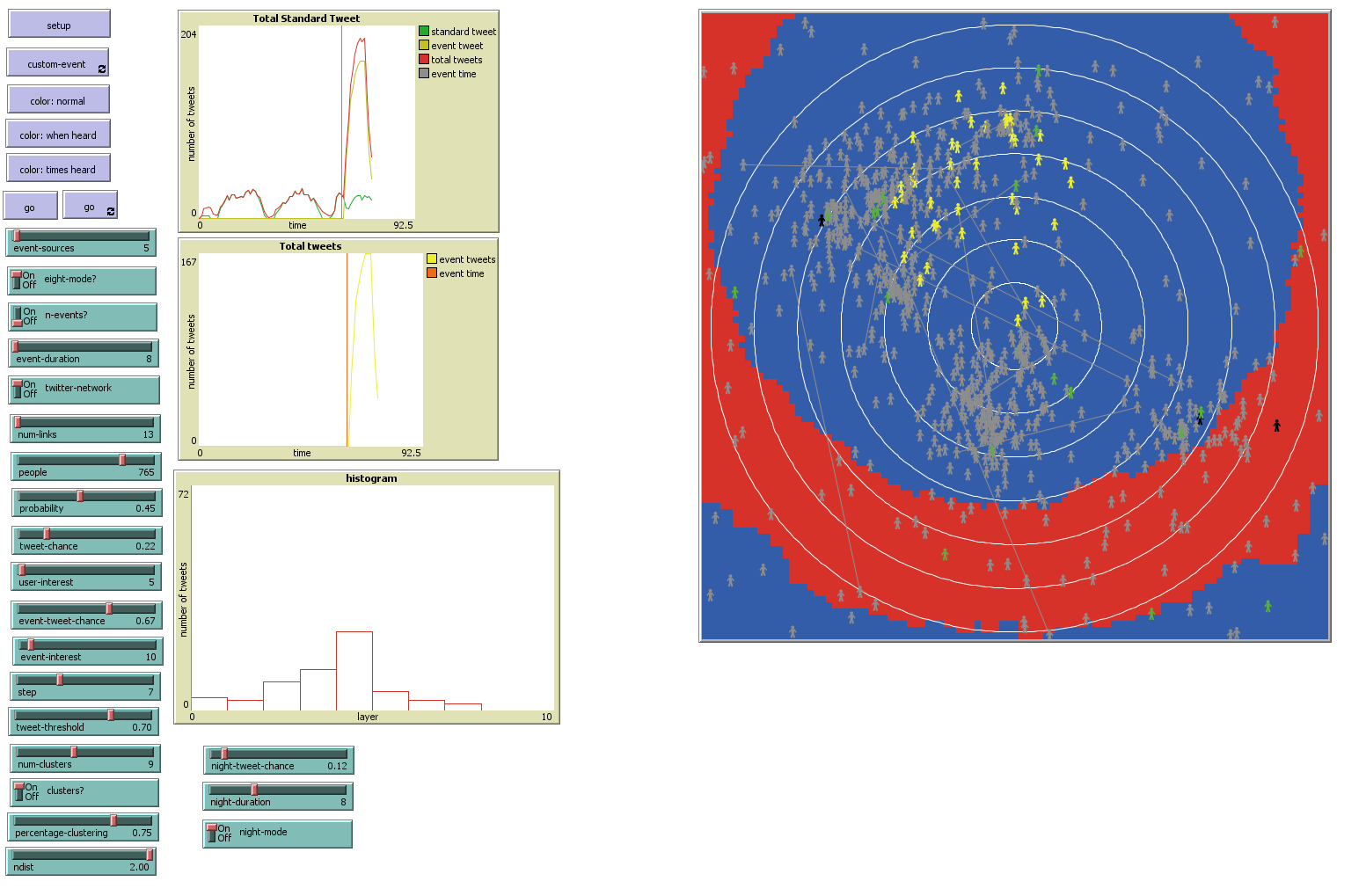}
  \caption{An example of TBAM's interface: the left side shows the parameters and the right side shows the simulation space}
  \label{fig:sim_world_comp}
\end{figure}

To generate data using TBAM, we assume that there are users with known locations distributed throughout the (synthetic) world. There is a reference sensor (we call it a social sensor) placed at a known location that counts the number of tweets at radial distances from itself. In Figure \ref{fig:sim_world_comp} the sensor is placed at the origin $(0,0)$. We believe that counting the number of tweets can give us reasonable information about microblogging behavior and we use this rather than focusing on the messages within the tweets (which need further processing). The changes in the number of tweets can be used for event detection. For example, previous work has shown that peaks in a time series of the number of tweets is an indication of an event \parencite{Mehdi2020Sem}. We assume the spreading of information about an event is analogous to rumor-spreading \parencite{jin2013epidemiological}. The rumor spreading model assumes that information about an event spreads out gradually similar to the ripple effect when a stone is dropped in a puddle of water. 

\begin{figure}[htp]
  \centering
  \subfigure[Plot of VIRG Twitter Data vs TBAM generated data]{\includegraphics[width=0.45\textwidth]{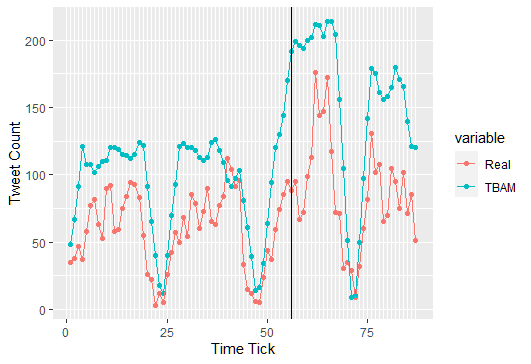}\label{fig:virg_abm}}\hfill
  \subfigure[Plot of STEM Twitter Data vs TBAM generated data]{\includegraphics[width=0.45\textwidth]{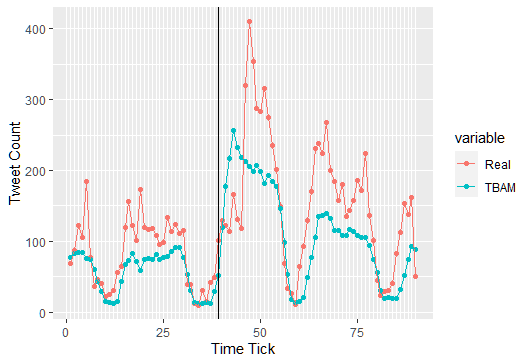}\label{fig:stem_abm}}\hfill
  \subfigure[Plot of Garlic Festival Twitter Data vs TBAM generated data]{\includegraphics[width=0.45\textwidth]{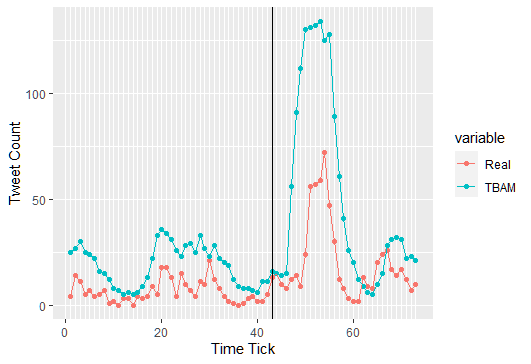}\label{fig:gar_abm}}
  \caption{Comparison of Real Twitter data with TBAM generated data}
\end{figure}

Figure \ref{fig:sim_world_comp} also shows different parameters like the probability that a user will tweet, distance and time from event, significance of an event, etc. By changing these parameters we have more control over the delivery slate. This allows generation of synthetic data that can match different scenarios. Finally, we validate our generated synthetic data by comparing it with data obtained from Twitter. Figures \ref{fig:virg_abm}, \ref{fig:stem_abm} and \ref{fig:gar_abm} show the comparison between the generated data using TBAM and the data obtained by scraping Twitter around three events. We describe the data sets later. 

In summary, our goals and contributions in this paper are as follows: 

\textbf{Formulation and Algorithm:}  We propose a methodology called Twitter Behavior Agent-Based Model (TBAM) and build a simulation using Netlogo to generate data using agent based modeling. Our model is able to identify the major parameters that affect the microblogging behavior in users.

\textbf{Accuracy:} Based on our results, we are accurately able to generate data that are statistically significant to real data as seen in Figures \ref{fig:virg_abm}, \ref{fig:stem_abm} and \ref{fig:gar_abm}.

\section{Literature Review} \label{lit_rev}

There are numerous surveys that focus on event detection and how humans behave when an event occurs \parencite{steiger2015advanced,atefeh2015survey,cordeiro2016online,garg2016review, imran2015processing, ajao2015survey,hasan2018survey,ozdikis2017survey,zheng2018survey}. It is believed that whenever an event occurs, there will be a change in the user behavior which will be reflected in the change in the microblogging activity. These papers have also mentioned how unreliable Twitter data is for event detection and that the data require considerable pre-processing before they can be used.

A part of the enrichment process for underdeveloped data is to use data generated through models. Generated data has been used in prior literature for \textit{data augmentation} and \textit{data imputation}. Data augmentation and imputation are relatively recently developed techniques. Data augmentation has been used in previous literature for image (e.g., facial data augmentation \parencite{wang2020survey}), speech and natural language processing (NLP) \parencite{dai2020analysis} and time-series data to reduce over-fitting \parencite{shorten2019survey,wen2020time}. Augmentation increases the size of the training data set by geometric and color transformations and deep learning techniques like Generative Adversarial Networks (GAN). Augmentation also alleviates the issue of class imbalance, which is a data set with skewed majority to minority sample ratios \parencite{shorten2019survey}. The effect of different augmentation techniques on time-series data was evaluated in \citeauthor{iwana2020empirical} (\citeyear{iwana2020empirical}) where there is also a guide for researchers and developers to help select the appropriate data augmentation method for their applications. Generative adversarial networks (GAN) was one of the popular methods used to generate synthetic images in the medical domain \parencite{bowles2018gan,frid2018gan,han2018gan}. These works generated images of CT images of liver lesions  \parencite{bowles2018gan,frid2018gan} and MR images \parencite{han2018gan} which were very close in comparison to the real data. Similarly, cycle-Consistent Generative Adversarial Networks (CycleGANs) were proposed as an image classification method to detect floods using images found in social media \parencite{pouyanfar2019unconstrained}. An agent-based model simulator called \textit{paysim} was created to simulate mobile money transaction and to create a synthetic data that is similar to the original data set \parencite{paysim}.

Data imputation is the task of estimating missing values in a data set. Data imputation is usually done to find missing values in traffic data arising from sensor damage, malfunction, or transmission errors, etc. using low-rank matrix decomposition. Most work on data imputation has focused on using GANs for data imputation by slightly varying its structure or the loss function \parencite{kim2020survey}. Two of the prominent works that have used GAN as a method for finding missing values in time-series data are found in \citeauthor{luo2018multivariate} (\citeyear{luo2018multivariate}) and \citeauthor{yoon2018gain} (\citeyear{yoon2018gain}).

However, there is little literature that has addressed the issue of using generated data to understand user microblogging behavior. The research studies have used Agent-Based modelling (ABM) to study information diffusion but none of the works focus on how user tweeting behavior changes when an event occurs. In \citeauthor{cui2013empirical} (\citeyear{cui2013empirical}) ABM was used to investigate how information was spread during the 2011 Wenzhou train crash through the Sina Weibo. They use the ABM framework to compare information diffusion through word-of-mouth and mass media and to determine which is a more significant means of spreading information when it comes to social media. ABM has been used to create an information propagation model to study how retweeting occurs \parencite{xiong2012information,pezzoni2013retweet}. In \citeauthor{pezzoni2013retweet} (\citeyear{pezzoni2013retweet}) a retweeting model was created based on two main parameters: the influence of the user (number of followers a user has) and the time at which the tweet was received. In \citeauthor{xiong2012information} (\citeyear{xiong2012information}) the retweeting model was based on the susceptible-infected-refractory (SIR) model. Similarly in \citeauthor{gatti2013large} (\citeyear{gatti2013large}), ABM was used to study user behavior in a social network. The model was created to predict the sentiment of users and whether they choose to forward, reply or do nothing about a topic.

\section{Background} \label{prelim}

In this section we look at how agent-based modeling and the Twitter network works. We design our agent-based model based on these definitions.

\subsection{Agent-Based Modeling} \label{abm}

ABM has been used in many different fields for analysis and understanding of the real world like biology, chemistry, cyber-security, social and economic modeling, etc. \parencite{allan2010survey}. Agent-based models (ABM) \parencite{wilensky2015introduction} have entities or \textit{agents}.  An agent can be an individual or an object that has specific properties and actions. The agents may move around in a two dimensional grid called the \textit{world}. The interactions between the agents can be quite complex but can be defined according to a set of rules. An agent can be autonomous, flexible, adaptable, and self-learning \parencite{macal2005tutorial}. There may be other models that can simulate scenarios pertaining to human social behavior and social media information dispersion (like system dynamics). However, ABM is able to better represent complex and heterogeneous interactions \parencite{wilensky2015introduction,macal2005tutorial} which makes it suitable for creating our model. Further, the microblogging behavior of one human may be considered actions of an agent, that may be influenced by what the agent sees in the environment and the actions of other agents. We believe that this is best modeled using ABM.

\subsection{Twitter as a Microblogging Service} \label{twit}

Twitter is currently the fastest growing (and by far the most popular) microblogging service with a lot of research being done on the generated content to understand human behavior \parencite{atefeh2015survey}. A Twitter user can send out a standard Twitter message called a \textit{tweet} about a specific topic and can contain a short text, links, or images. Messages can be grouped together based on their topic or group by use of their \textit{hastags}. Tweets can also be forwarded by other users and they are called \textit{retweet}s. A retweet can only be received if a user is in the same network as the user who originally sent out the tweet. The tweets for research and analysis can be obtained from the Twitter API\footnote{https://dev.twitter.com/overview/api}. The Twitter API can provide past tweets as well as stream tweets. The Twitter API only allows tweets to be collected over the past two weeks and has restrictions in terms of the amount of number of tweets that can be collected.

\section{Twitter Behavior Agent-Based Model (TABM) Description} \label{method}

To generate synthetic data we developed the Twitter Behavior Agent-Based Model (TBAM) that uses Agent-Based Modelling (ABM) to simulate how users tweet and how microblogging behavior changes when an event occurs. In this section we provide a detailed overview of the model and the different parameters used to generate TBAM data.

\subsection{TBAM Design}

The scope of TBAM is to simulate user behavior when an event occurs, more specifically to investigate the change in number of tweets as time and distance from event changes. For our model we consider "local events" \parencite{ozdikis2017survey}, i.e., events restricted to a certain region. Our model simulates microblogging behavior of people similar to what may happen within a city or a few small neighborhoods and helps us examine how people's microblogging behavior changes when they have close spatial and temporal proximity to an event.

We use \textit{Netlogo} \parencite{Netlogo} to create our agent based model. In Netlogo there are four types of agents: turtles, patches, links and observer. The turtles are agents that move around in the \textit{world}. The world is sub-divided into smaller squares called \textit{patches} and each patch has a unique coordinate. Links are agents that connect two turtles. The observer observes the agents and their interactions. In Netlogo models, time passes in discrete steps called \textit{ticks}.

Figure \ref{fig:sim_world_comp} shows a snapshot of the synthetic world at a particular tick. In our model, the turtles are the Twitter users (people) who send out the tweets. A tweet can be a non-event related tweet (which is a standard or routine tweet indicated by green colored users), an event related tweet (indicated by users colored yellow) or tweets sent out during low Twitter activity (indication by users colored black). The patches represent the locations over which Twitter users lie and where an event can occur. Initially all patches are colored blue. Once an event occurs, the patches change color to red as the patches are influenced by the event. In a real world setting a patch could represent a geographical coordinate. A tick is a unit of time over which the total number of tweets are measured. Ticks could be in hours, minutes or seconds depending on the time granularity that is required. 

\subsection{TBAM Parameter Description} \label{tbam_desc}

In order to generate data that may accurately reflect real world settings we define different parameters. The parameters are summarized in Table \ref{tab:synth_exp2a} and Table \ref{tab:synth_exp2b}. In this section, we provide an overview of these parameters and explain how our model simulates microblogging behavior and event generation. The model is made of two phases. In the first phase, also called the \textit{setup} phase, the synthetic world settings are created in which the users will tweet. In the second phase, also called the \textit{simulation} phase, the users tweet and once an event occurs, their microblogging behavior changes. 

\begin{table*}[htp]
\caption{Fixed parameters for all TBAM data generation simulations}
\centering
\begin{tabular}{| p{.25\textwidth} | p{.55\textwidth} | p{.10\textwidth} |}
\hline
\textbf{Parameter} & \textbf{Description} & \textbf{Value} \\
\hline
\texttt{n-events?} & binary: chooses between one event or n-events & FALSE \\
\hline
\texttt{event-sources} & sets number of events. Only valid if n-events is true & 1 \\
\hline
\texttt{eight-mode?} & binary: chooses between spreading event to both diagonal and adjacent patches or only to adjacent patches & true \\
\hline
\texttt{twitter-network} & binary: chooses between Erdos-Renyii and random network & Erdos-Renyii \\
\hline
\texttt{num-links} & number of links in the random network & -NA- \\
\hline
\texttt{people} & number of people microblogging & 1000 \\
\hline
\texttt{probability} & probability of a link being created between two people in the Erdos-Renyii Network & 0.45 \\
\hline
\texttt{step} & distance between each layer & 7 \\
\hline
\texttt{num-clusters} & number of clusters of people & 9 \\
\hline
\texttt{cluster?} & binary: choose to cluster people (1) or distribute people uniformly (0) & 1 \\
\hline
\texttt{percentage-clustering} & percentage of people in clusters & 0.75 \\
\hline
\texttt{tweet-threshold} & used to generate random number that a user will tweet & 0.7 \\
\hline
\texttt{user-interest} & duration over which people remain interested about a tweet & 5 \\
\hline
\texttt{event-interest} & duration over which people remain interested about tweets related to an event & 5 \\
\hline
\texttt{night-mode} & to consider periodicity in tweets and separate users microblogging at day and night & True \\
\hline
\end{tabular}
\label{tab:synth_exp2a}
\end{table*}

The setup phase begins with the creation of $N$ users. The users are randomly distributed throughout the world. Some of the users are clustered together. The number of clusters are defined by the parameter \texttt{num-clusters}. The parameter \texttt{cluster?} determines if the users are clustered or not and \texttt{percentage-clustering} determines what percentage of users are clustered together randomly in each of the clusters. The setup phase also generates concentric circles around the central coordinate, i.e $(0,0)$. The central coordinate is the location of the sensor that counts the number of tweets. Each consequent circle increases its radius by the parameter \texttt{step}. These circles aim to provide a visual understanding of how tweets change with changing distance.

To simulate a Twitter network, some users are linked together with bi-directional links. The links are generated using \textit{Erdos-Renyi} model which has been used in previous literature to study social networks \parencite{erdHos1960evolution}. In the Erdos-Renyi model, each link between a user has a fixed probability of being present and being absent independent of the links in a network.  The parameter \texttt{probability} can be varied to change the probability and create networks with a lot of links or with very few links between users. There is also an option for generating a network with random links between users. The \texttt{num-links} is a parameter specific to random networks which randomly creates \texttt{num-links} number of links between different users. The \texttt{twitter-network} can be set to true or false to choose between Erdos-Renyi or random model to generate the network.

In the simulation phase at each tick the total number of tweets sent out are counted. The total number of tweets are the sum of standard tweets, event related tweets, and tweets sent out during low Twitter activity. It should be noted that we are able to separate these in TBAM but it may not be possible to do so with scraped Twitter data. At each tick, a random number $z_i$ is generated for each user. The random number is used to create the random conditions where users may not choose to tweet at a specific time. Since it is hard to determine these random conditions, we assume that $z_i$ is normally distributed random number of mean \texttt{tweet-threshold} and variance $0.2$. The scraped twitter data also follows a rough normal distribution which is why we chose the random conditions to be normally distributed. Before an event occurs a user will only send out a routine tweet. A user will only tweet if $z_i < \texttt{tweet-chance}$ where $z_i$ is a random number generated for user $i$ and \texttt{tweet-chance} is probability of sending out a standard tweet (from Table \ref{tab:synth_exp2b}).

\begin{table*}
\caption{Variable parameters for different TBAM data generation simulations}
\centering
\resizebox{\textwidth}{!}{\begin{tabular}{| p{.20\textwidth} | p{.40\textwidth} | p{.08\textwidth} | p{.08\textwidth} | p{.08\textwidth} |}
\hline
\textbf{Parameter} & \textbf{Description} & \textbf{VIRG} & \textbf{STEM} & \textbf{GAR} \\
\hline
\texttt{tweet-chance} & probability of a person sending out a tweet & 0.33 & 0.29 & 0.22 \\
\hline
\texttt{event-duration} & length of time that event remains active & 31 & 48 & 8 \\
\hline
\texttt{event-tweet-chance} & probability of a person sending out a tweet about an event & 0.49 & 0.55 & 0.67 \\
\hline
\texttt{night-tweet-chance} & probability of sending out tweets during low Twitter activity & 0.17 & 0.16 & 0.12 \\
\hline
\texttt{night-duration} & the duration of low Twitter activity & 8 & 8 & 8 \\
\hline
\texttt{ndist} & scaling factor effecting decay for $q_i$ & 0.07 & 0.12 & 2 \\
\hline
\end{tabular}}
\label{tab:synth_exp2b}
\end{table*}

At a specific time (tick) and location (patch) the event occurs and with each tick spreads across the world. The rumor spreading model has been used as a basis to simulate spreading of an event influence (or information) \parencite{RumorMill}. There may be other models that can be used to simulate spreading of an event, like the susceptible-infected-refractory (SIR) model \parencite{xiong2012information}. In a manner similar to the rumor spreading model, immediately after an event occurs, the event influence starts spreading to all of the neighboring patches (shown by the red colored patches in Figure \ref{fig:sim_world_comp}). However, the rate at which the event influence spreads to its adjacent neighbors may not be uniform and may vary with time. Initially, as soon as the event occurs, the event influence immediately spreads to all patches within a fixed radius. The influence then spreads to adjacent neighboring patches with decreasing rate as more time elapses. This assumption is based on the observations made from collected Twitter data that show a sharp rise in the counts immediately after an event. The parameter that affects the spreading of an event is \texttt{eightmode?}. Setting \texttt{eightmode?} to true causes the event to spread to its diagonals and its adjacent neighbours but setting the \texttt{eightmode?} to false causes the event to only spread to its adjacent neighbours. It should be noted that when \texttt{eightmode?} is true, then the event spreads outwards more quickly. 

Once an event occurs, a user can send out either an event related tweet or a routine tweet. A user will choose to send out a tweet about an event if $z_i < \frac{q_i}{(q_i + \texttt{tweet-chance})}$ where $z_i$ is a random number generated for user $i$ as described previously, $q_i$ is probability a user $i$ will tweet about an event and \texttt{tweet-chance} is from Table \ref{tab:synth_exp2b}. Once a user chooses to tweet about an event, then a user will only send out tweet about an event if they are on a patch where an event has spread to and $z_i < q_i$ where $z_i$ is a random number generated for user $i$ and $q_i$ is probability a user $i$ will tweet about an event. There are multiple methods of determining $q_i$. The value of $q_i$ could be fixed or vary with time and distance from event. For the model, we employ a hybrid approach. In the immediate vicinity of the event $q_i = \texttt{event-tweet-chance}$ where \texttt{event-tweet-chance} is one of the parameters from Table \ref{tab:synth_exp2b}. But as the distance and time from event increase, $q_i$ decreases according to Equation \ref{eq:01}.

\begin{equation}
    q_i = event-tweet-chance*[ (t-t_{event})^{-\texttt{ndist}/\alpha} * (d_{event})^{-\texttt{ndist}/\beta} ] 
    \label{eq:01}
\end{equation}

\begin{figure}[htp]
  \centering
  \includegraphics[width=8cm,height=6cm]{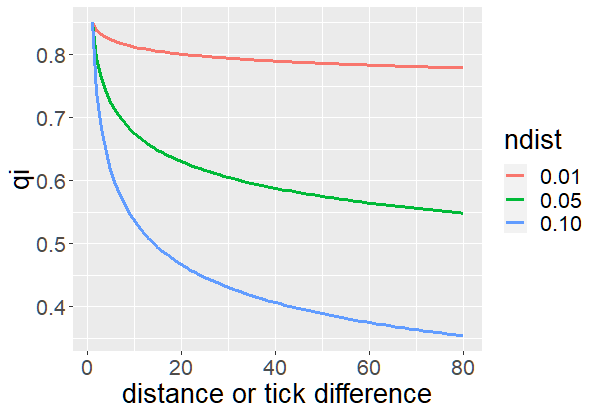}
  \caption{Changing $q_i$ with changing distance or ticks (with $\alpha=1$ and $\beta=20$)}
  \label{qi}
\end{figure}

The variable $t$ is the current time tick measured after the event occurs, $t_{event}$ is the time tick at which event occurred, $d_{event}$ is the distance of the user from the event, $\texttt{ndist}/\alpha$ and $\texttt{ndist}/\beta$ are scaling factors. A high \texttt{ndist} value means that \texttt{event-tweet-chance} decays less rapidly with changing time and distance. For our model we keep $\alpha$ fixed at $1$ and $\beta$ fixed at $20$. Since we are considering local events and users would generally be in close proximity to the event, the decay of \texttt{event-tweet-chance} with distance should be less rapid than decay due to time. Hence, we choose a larger value of $\beta$ than for $\alpha$. It should be noted that there are many different functions that could be used to simulate the decay of probability of microblogging about an event. But previous literature \parencite{pezzoni2013retweet,sakaki2010earthquake} have considered exponential distribution for tweets which can also be observed from the collected twitter data. Hence, we choose a simple exponential function to change how the probability of microblogging about an event decays with time and distance in our simulations. Figure \ref{qi} is a plot of the function showing how $q_i$ changes as distance and time from event changes when $\alpha=1$ and $\beta=20$.

During the simulation phase, users can also send out retweets. A user will send out a standard retweet if $z_i < \texttt{tweet-chance}$ AND there is a link with another user who has sent out a standard tweet. Consequently, a user will only send out an event related retweet, if $z_i < q_i$ and there is a link with another user who has sent out an event related tweet. We define the parameters \texttt{event-interest} and \texttt{user-interest} as the tick duration over which users will keep on talking about an event or a routine tweet.  These parameters quantify the importance of standard or event related tweets and the higher these parameters are, the larger will be the number of retweets sent out. For our simulations, we keep these values constant for the simulation duration. 

The event ends when \texttt{event-duration} ticks have elapsed. Once the event ends, just like the rumor spreading model, the patches loose the influence of the the events. No new tweets relating to an event are generated and the event related tweets gradually decrease until they eventually stop. A higher \texttt{event-duration} value signifies that users will continue to generate new tweets about an event for longer periods of time.

There is also the option of choosing multiple events which is done by setting \texttt{n-events?} true. If there are multiple events then \texttt{event-sources} sets the number of event sources. For this paper, we use one event source. In short, \texttt{event-duration}, \texttt{event-tweet-chance} and \texttt{event-interest} determine how significant an event is. If these values are set high then it indicates an event that has a high impact on users' lives and they will tweet and retweet more about the event and remain interested in the event for longer duration. These parameters can be changed to incorporate different types of events. 

The data collected from Twitter reveals periods of time with very few tweets being sent out. To incorporate such behavior we introduce the parameter \texttt{night-mode} that enables or disables consideration of time when there is low Twitter activity. If \texttt{night-mode} is enabled, then there are two parameters that affect the low Twitter activity. One parameter \texttt{night-duration} effects how long the low activity period lasts. The other parameter \texttt{night-tweet-chance} is a measure of the probability of a user microblogging during the low Twitter time period. A user sends out tweets during this time only when $z_i < \texttt{night-tweet-chance}$. Usually  \texttt{night-tweet-chance} would be less than \texttt{tweet-chance} which in turn is usually less than \texttt{event-tweet-chance}.  

\section{TBAM validation} \label{twitter_real}

For our comparison we used three data sets, referred to as VIRG, STEM and GAR, collected directly from the Twitter API using the 'TwitteR' package in R \parencite{TwitteR}. The social sensor (reference point) coordinate was 2 miles from the event along the y-axis. It shows the number of tweets within a 2.8 mile radius changing with time. The data are related to three events and are summarized in Table \ref{tab:real_dataset}. Figures \ref{fig:virg_abm}, \ref{fig:stem_abm} and \ref{fig:gar_abm} represent the plots of the data. \textit{Real} indicates data obtained from Twitter and \textit{TBAM} indicates data generated through TBAM. Each tick represents the number of tweets sent out in an hour. The occurrence of an event is indicated by the vertical line. From the plots it can be clearly seen that after an event occurs, there is a sharp rise in the number of tweets which is similar to the real data.

\begin{table}
  \caption{Summary of Real Data} 
  \centering
  \begin{tabular}{| p{3cm} | p{1.5cm} | p{2cm} | p{3.5cm} | p{3.1cm} |} 
  \hline %inserts double horizontal lines
  \textbf{Event Name} & \textbf{Reference Name} & \textbf{Event Date} & \textbf{Event Location (latitude, longitude)} & \textbf{Social Sensor Location (latitude, longitude)} \\ 
  \hline
  Virginia Beach Shootings & VIRG & 05-31-2019 4:44pm & 36.7509,-76.0575 & 36.77974,-76.05750 \\
  \hline
  STEM School Shootings & STEM & 05-07-2019 1:53pm &  39.556,-104.9979  & 39.58482,-104.99790 \\  
  \hline
  Garlic Festival Shootings & GAR & 07-28-2019 5:40pm & 36.997778,-121.585278 & 37.02661,-121.58528 \\
  \hline
  \end{tabular}
  \label{tab:real_dataset} 
\end{table} 
 
To generate data using TBAM, we use the parameter values described in Table \ref{tab:synth_exp2a} and Table \ref{tab:synth_exp2b}. Table \ref{tab:synth_exp2a} shows the parameters that are kept fixed for all the simulations. Table \ref{tab:synth_exp2b} shows the parameters that are changed according to the Twitter data set they are meant to match. The parameters in Table \ref{tab:synth_exp2b} were estimated by inspection of the Twitter data. Table \ref{tab:synth_exp2c} summarizes how we estimated the different parameters from the Twitter data. The \texttt{event-duration} was estimated as the duration over which the number of tweets sent after the event were higher than tweets sent before the event. For example, in Figures \ref{fig:stem_abm} the number of tweets in the \textit{Real} data return to the value before the event after 48 ticks, hence, the TBAM \texttt{event-duration} parameter was set to 48 to generate the data in that figure.

Similarly, from the real data we calculate the values for \texttt{$tweets_{night}$}, \texttt{$tweets_{pre-event}$} and \texttt{$tweets_{post-event}$}. Then using these values we estimate the probabilities for the TBAM parameters of \texttt{tweet-chance}, \texttt{event-tweet-chance} and \texttt{night-tweet-chance} which are then used to generate TBAM data.

The heuristic analysis of the data from Twitter reveals how the different parameters vary for different areas and events. The difference in these parameters could be due to the difference in demographics, Twitter usage, and density of the Twitter network. For all three data sets we considered a similar event. VIRG and STEM had roughly similar parameters but GAR has very different parameters. This is because GAR refers to two events combined as one. GAR event is different from the other two events as it started off as a different event which was a festival but ended up as a shooting event. As a result there were more Twitter users compared to usual days and the parameters \texttt{event-tweet-chance}, \texttt{event-duration} and \texttt{ndist} are significantly different than the other two events. The parameter of \texttt{event-duration} is significantly shorter due to Twitter users leaving the event location and hence, the scaling factor is high to account for the high outflow of Twitter users. 

\begin{table}[htp]
\centering
\begin{tabular}{| p{.35\textwidth} | p{.55\textwidth} |}
\hline
\textbf{Parameter} & \textbf{Description} \\
\hline
\texttt{$tweets_{night}$} & mean number of tweets sent in low Twitter activity hours \\
\hline
\texttt{$tweets_{pre-event}$} & mean number of tweets sent before the event (excluding low Twitter activity tweets) \\
\hline
\texttt{$tweets_{post-event}$} & mean number of tweets sent after the event (excluding low Twitter activity hour tweets) \\
\hline
\hline
\texttt{event-duration} & duration over which number of tweets sent after event are remain higher than number of tweets sent before event \\
\hline
\texttt{tweet-chance} &  \Large $\frac{tweets_{pre-event}}{(tweets_{night})+tweets_{pre-event}+tweets_{post-event}}$ \\ 
\hline
\texttt{event-tweet-chance} & \Large $\frac{tweets_{post-event}}{(tweets_{night})+tweets_{pre-event}+tweets_{post-event}}$ \\
\hline
\texttt{night-tweet-chance} & \Large $\frac{tweets_{night}}{(tweets_{night})+tweets_{pre-event}+tweets_{post-event}}$ \\
\hline
\end{tabular}
\caption{Determining the probabilities from Twitter data for TBAM}
\label{tab:synth_exp2c}
\end{table}

 \subsection{Model Validation}

In order to measure the accuracy of the TBAM generated data compared to the data collected from Twitter, we use the \textit{cross-correlation function (\textit{ccf})} \parencite{shumway2000time}. The cross-correlation function between two time series $x_t$ and $y_t$ is given by:
\begin{equation}
    \rho_{xy}(s,t) = \frac{\gamma_{xy}(s,t)}{\sqrt{\gamma_{x}(s,s)\gamma_{y}(t,t)}} 
\end{equation}
where
\begin{equation}
    \gamma_{xy}(s,t) = cov(x_{s},y_{t}) = E[(x_{s}-\mu_{xs})(y_{s}-\mu_{yt})]
\end{equation}
$\mu_{xs}$ is the mean of time series $x_s$ and $\mu_{yt}$ is the mean of time series $y_t$.

The cross-correlation measures the dependence between two points on different time series observed at different times. In other words, \textit{ccf} measures the linear predictability of the series at time $s$, say $x_s$, using only
the value $y_t$. In our TBAM data we are trying to see if the trends and patterns in the original data match. Hence, \textit{ccf} will be a suitable metric and give an overview of the statistical significance between the two data sets.

\begin{figure}[htp]
  \centering
  \subfigure[Cross Correlation between VIRG Twitter Data and TBAM data]{\includegraphics[width=0.30\textwidth]{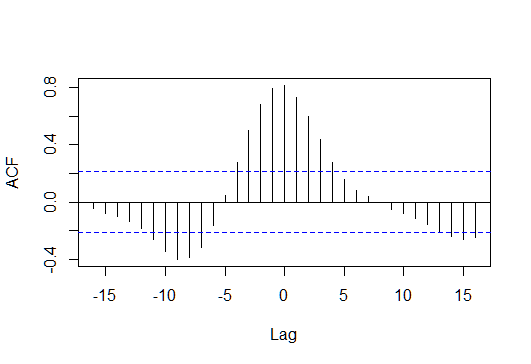}\label{fig:ccf_virg1}}\hfill
  \subfigure[Cross Correlation between STEM Twitter Data and TBAM data]{\includegraphics[width=0.30\textwidth]{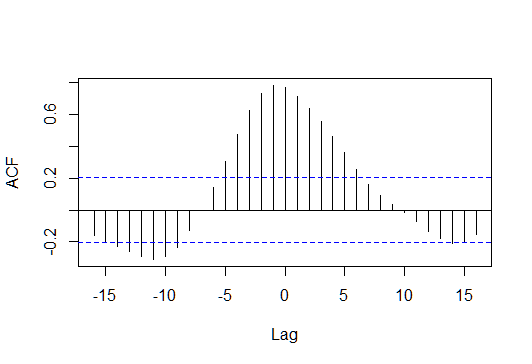}\label{fig:ccf_stem1}}\hfill
  \subfigure[Cross Correlation between GAR Twitter Data and TBAM data]{\includegraphics[width=0.30\textwidth]{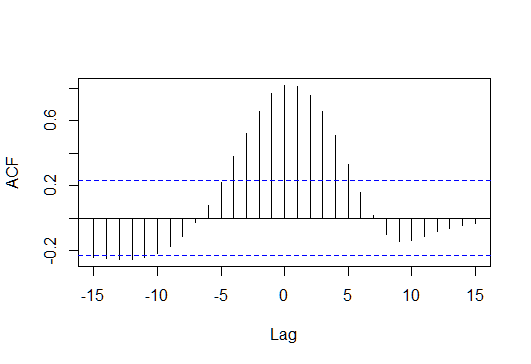}\label{fig:ccf_gar1}}
  \caption{CCF of Real Twitter data with TBAM generated data}
\end{figure}

Figures \ref{fig:ccf_virg1}, \ref{fig:ccf_stem1} and \ref{fig:ccf_gar1} shows the \textit{ccf} between VIRG, STEM and GAR data sets and the TBAM generated data respectively. It can be seen that the TBAM generated data and the real data have a very similar pattern as the \textit{ccf} is higher than the threshold for most lags. Another important observation is that the correlation is also high at $lag=0$. High correlation at $lag=0$ indicates a strong statistical significance and shows that the patterns in both the data match very closely. This also means that any events found in the real data could also be found in the same position in the TBAM data.

\begin{figure}[htp]
  \centering
  \subfigure[Cross Correlation between VIRG Twitter Data and Uniform Random Data]{\includegraphics[width=0.30\textwidth]{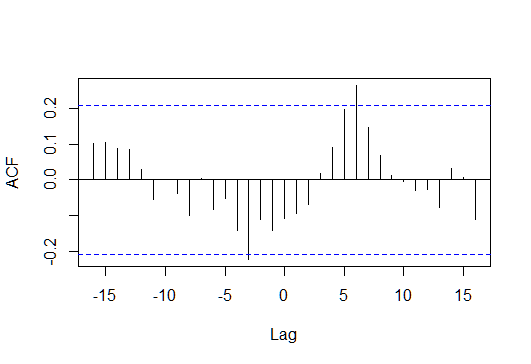}\label{fig:ccf_virg2}}\hfill
  \subfigure[Cross Correlation between STEM Twitter Data and Uniform Random Data]{\includegraphics[width=0.30\textwidth]{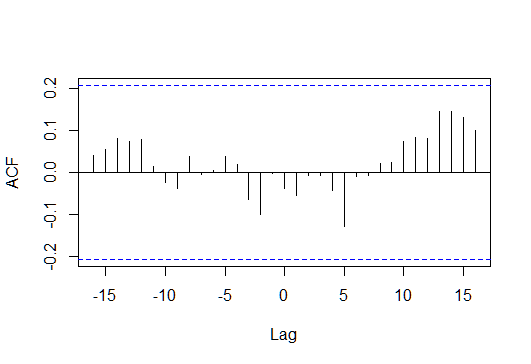}\label{fig:ccf_stem2}}\hfill
  \subfigure[Cross Correlation between GAR Twitter Data and Uniform Random Data]{\includegraphics[width=0.30\textwidth]{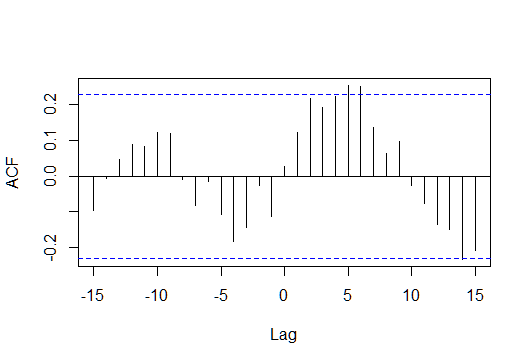}\label{fig:ccf_gar2}}
  \caption{CCF of Real Twitter data with randomly generated data}
\end{figure}

For comparison, to show that our method does generate reasonably accurate data, we measure the \textit{ccf} between data from Twitter and uniformly randomly generated data. The random numbers were generated between 1 and the maximum value in the number of tweets of the specific data set. Figures \ref{fig:ccf_virg2}, \ref{fig:ccf_stem2} and \ref{fig:ccf_gar2} shows the \textit{ccf} between VIRG, STEM and GAR data sets and uniformly randomly generated data. It is clearly seen from the plots that there is very low correlation at all lags. This shows that the data generated using TBAM is significantly better. 

As shown above, TBAM reasonably reproduces microblogging behavior. Unlike real tweets, here we can (a) identify, control, and tune parameters which impact the tweet counts and microblogging behavioral patterns (b) separate event related tweets from standard tweets without looking at semantics (c) create aggregates in space or time (which we have not done here, but it is straightforward) which can reflect real world data scraping and provides different space and time granularity (d) intentionally, randomly, or using other models, introduce gaps or noise (e) add additional demographics - groups that tweet more or less (f) add system-wide or regional variations - all in a controlled manner. In this way, there is more control over the delivery slate and reliability of the microblogging data.

\section{Conclusion} \label{conc}

The result from our analysis indicates that the data generated using TBAM can be used to complement and possibly enrich the underdeveloped real-world data. The main application of our data generation model is in detecting and localizing events using crowd-sourced “social sensors” whose aggregate counts of event alerts are collected at reference sensors. The generated data can be used to enrich the data collected from Twitter. This can potentially improve event detection accuracy that may enable applications like first response during minor, yet consequential disasters such as road closures. Our model may also be used for data augmentation and imputation (estimation), for generating training data for machine learning models, testing event detection models and determining parameters that determine tweeting behavior of people.

An important method in event detection are machine learning methods (like long short-term memory (LSTM), classification, etc.) \parencite{atefeh2015survey}. The machine learning methods can be used to find event patterns in number of tweets. In case of event detection there might be the problem of class imbalance as there might be a single event signature in a long sequence of tweets. This can cause errors in event pattern detection when using machine learning methods. There might also be the problem of insufficient tweets to train the machine learning methods. In such cases the TBAM generated data can be used for data augmentation by creating additional synthetic data that can be used to train the machine learning methods.

Other previous works, such as \citeauthor{Mehdi2020Sem} (\citeyear{Mehdi2020Sem}) and \citeauthor{comito2017peak} (\citeyear{comito2017peak}), have considered peaks as indication of an event and have created models to find events using peaks. In fact a quick visual analysis does show that peaks in TBAM generated data and the data collected from Twitter have peaks in very close proximity to each other. We will perform detailed comparison of the peaks in future work. Nevertheless it does show that the data generated using TBAM can be used in lieu of Twitter data to find peaks and detect events. 

Another application of the model could be to create a data base that would list different probabilities for different locations and different event-types. Our model quantified and identified different parameters that effect the microblogging behavior of users. These parameters could then be used to identify different events.

In our model we made simple assumptions for the parameters, some of which we estimated from the real data and some parameters, which could not be determined from the real data, were randomly generated. For example, the distribution of the people and their clusters was random which may not be the case in real world. Furthermore, each user is assigned a random number $z_i$ to reflect the inclination of a user to tweet. As a result, the magnitude of the number of tweets in the TBAM data and real data are different. As part of future work, we aim to design a method to better estimate these parameters. Different event spreading techniques and equations for decay of probability of microblogging about an event can be applied and tested. One way to better estimate the parameters would be to create an optimization problem that maximizes \textit{ccf} or use past Twitter data. But since Twitter data is underdeveloped, we could generate data using alternative methods like generative adversarial networks (GAN) and use GAN generated data to learn TBAM parameters. In this way both TBAM and GAN could work together to further enrich the data. Alternatively, we could also compare TBAM and GAN data generation techniques.

%\bibliographystyle{plainnat}
%\bibliography{datafusion}

\printbibliography[heading=bibliography]

\end{document}